\documentclass[conference]{IEEEtran}
\IEEEoverridecommandlockouts
\usepackage{cite}
\usepackage{amsmath,amssymb,amsfonts}
\usepackage{algorithmic}
\usepackage{graphicx}
\usepackage{textcomp}
\usepackage{xcolor}
\usepackage{subfigure}
\usepackage{wrapfig}
\usepackage{amsmath}
\usepackage{subfig}
\usepackage{url}
\usepackage{manyfoot}%
\usepackage[utf8]{inputenc}

\def\BibTeX{{\rm B\kern-.05em{\sc i\kern-.025em b}\kern-.08em
    T\kern-.1667em\lower.7ex\hbox{E}\kern-.125emX}}
\begin{document}

\title{Quantitative Analysis of Forecasting Models:\\ In the Aspect of Online Political Bias
}

\author{\IEEEauthorblockN{Srinath Sai Tripuraneni, Sadia Kamal, Arunkumar Bagavathi}
\IEEEauthorblockA{\textit{Department of Computer Science} \\
\textit{Oklahoma State University}\\
%Stillwater, Oklahoma, USA \\
stripur@okstate.edu, sadia.kamal@okstate.edu, abagava@okstate.edu}
}

\maketitle

\IEEEpeerreviewmaketitle

\begin{abstract}
Understanding and mitigating political bias in online social media platforms are crucial tasks to combat misinformation and echo chamber effects. However, characterizing political bias temporally using computational methods presents challenges due to the high frequency of noise in social media datasets. While existing research has explored various approaches to political bias characterization, the ability to forecast political bias and anticipate how political conversations might evolve in the near future has not been extensively studied. In this paper, we propose a heuristic approach to classify social media posts into five distinct political leaning categories. Since there is a lack of prior work on forecasting political bias, we conduct an in-depth analysis of existing baseline models to identify which model best fits to forecast political leaning time series. Our approach involves utilizing existing time series forecasting models on two social media datasets with different political ideologies, specifically Twitter and Gab. Through our experiments and analyses, we seek to shed light on the challenges and opportunities in forecasting political bias in social media platforms. 
%Ultimately, our work aims to pave the way for developing more effective strategies to mitigate the negative impact of political bias in the digital realm.
\end{abstract}


\section{Introduction}
% (\textit{Example:} more than 80\% of Americans receive updates about COVID-19 from social media websites~\cite{casero2020impact}). Several research studies claim news media bias and polarized user communities on social media 
News media houses have endured through time to disseminate political news to the people while also influencing their political perceptions. The immense growth of online social media has a significant effect on how news is being consumed in recent years, giving them resources to seed disinformation and fake news on the course of accelerating the information dissemination process~\cite{vicario2019polarization}. The causalities of social media polarization can be computationally characterized with three aspects: \emph{time-based}~\cite{garimella2017long}, \emph{topic-based}~\cite{vicario2019polarization} and \emph{user-based}~\cite{bonchi2019discovering}.  \emph{Time-based} approaches qualitatively analyze dynamics of politically biased topics in online forums, \emph{Topic-based} approaches characterize polarization with linguistic queues on content-level details (entities, topics, etc.) and how social media communities react with the multitude of opinions to such contents. \emph{User-based} approaches formulate polarization with user communities in the social network and how topics help to divide communities in the network. 
%Most often both approaches are mutually inclusive, where it is unrealistic to study polarization without one or the other.

% we give a politically biased time series as input and the forecasting model predicts the next step or any future developments as output
%By accurately predicting how political sentiments unfold over time, we
In this work, we explore a novel research endeavor focused on forecasting of political bias in two social media platforms. We formulate this as a time series forecasting problem where the objective is to capture correlation between political bias and information evolving patterns. Such temporal forecasting can give insights to analysts on the formation of ideological clusters and the dissemination of biased information on social media platforms. Notably, prior research in forecasting political bias data is limited, making our exploration a pioneering effort in this domain. We leverage existing time series forecasting models to evaluate their suitability for this task. By analyzing these models' performance in forecasting political leaning time series, we aim to uncover their strengths and limitations in capturing the temporal dynamics of political bias. Overall, we have a \emph{two-fold} contributions in this paper:
%  Since each social media platform has its own user participation level, popularity, and political ideology, we experiment with the time series forecasting problem on two social media datasets, \emph{Twitter} and \emph{Gab}, collected during the same timeframe.
% This can ultimately aid in devising effective strategies to mitigate misinformation and echo chamber effects.

\begin{enumerate}
    \item \textbf{Contribution-1:} We propose a new problem of forecasting the political bias on online social media posts. Such forecasting is crucial for understanding the social media's political standpoint on any given topic or event
    \item \textbf{Contribution-2:} We experiment with various time series forecasting models to quantify the trends in different political biases of two social media forums that have different user participation, popularity, and political ideology: Twitter and Gab. 
\end{enumerate}

% \begin{figure*}[htp]
%   \centering
%   \subfigure[Average political leaned tweets vs months]
%   {\includegraphics[scale=0.19]{1.png}}\quad
%   \subfigure[Total political leaned tweet likes per month vs months]
%   {\includegraphics[scale=0.19]{2.png}}
%   \caption{Frequency of tweets and tweet likes for each political leaning}
%   \label{fig:tweets_like}
% \end{figure*}

% \begin{figure*}[htp]
%   \centering
%   \subfigure[Average political leaned GAB posts vs months]
%   {\includegraphics[scale=0.21]{3.png}}\quad
%   \subfigure[Frequency of GAB posts likes for each political leaning]
%   {\includegraphics[scale=0.21]{4.png}}
%   \caption{Frequency of Gab posts and  likes for each political leaning}
%   \label{fig:Gab_like}
% \end{figure*}

\section{Related Work}
%They play a vital role in impersonating public opinion on several issues like diseases~\cite{alafnan2020covid}, elections~\cite{beckers2020voice} etc. It is an established fact that the sharing of politically biased information has increased rapidly after the invention of Social media. Hence, predicting political leaning has been emerging from time to time with new research ideas to stop the bad consequences of sharing politically biased information.

% In this section, an exhaustive review of the existing related work is discussed focusing on how there was progress in the prediction of political leaning in various applications and how those methodologies are different from the methodology that was applied in this paper.

% \subsection{Using Content and hidden structuring of the tweets}\label{AA}
Many studies have focused on analyzing the content of tweets to predict the political inclination of individual users. Jiang et al.~\cite{jiang2023retweet} introduced an NLP model named retweet- BERT which utilizes retweet networks for prediction of political leaning of users.
Another study \cite{9680097} analyzed content of the tweets of the users to identify their corresponding political leaning. 
Efron et al. \cite{efron2006using} have used a probabilistic model to estimate the political orientation of documents.
% \subsection{Political leaning prediction limited to certain locations.}
Significant research has been done to predict political leaning which is on tweets limited to certain locations. Work in \cite{cardaioli2020predicting} tried to forecast whether users are more left- or right-oriented in different languages.
% Examining the writings of journalists who support the government as well as those who oppose it.
% Carrying out a relevance analysis to assess the significance of terms used within each category of journalists. Conducting sentiment analysis to gauge the emotional tone conveyed in the texts.
% Utilizing topic modeling techniques to examine the themes and subjects covered in the documents.
% Analyzing psycholinguistic indicators derived from the Linguistic Inquiry and Word Count (LIWC) system.Another study\cite{rahmati2023predicting} focuses on analyzing political beliefs by employing deep learning techniques. Specifically, it leverages the properties of LSTM  and CNN  with BERT. Nowadays there are many microblogging platforms that have a high chance of increasing politically leaned users to express their politically leaned views freely. So, some of the researchers started to explore how to predict political leaning in these platforms as well. Kitchener et al. \cite{kitchener2022predicting} have utilized the Reddit online discussion forum by applying a set of statistical learning approaches to predict individual political ideology.
Turkmen et al. \cite{turkmen2014political} implemented a Support Vector Machine and Random Forest Classifiers connected with a statistic-based feature selection to predict political tendency on a small selection of political communications.  Another study \cite{8531001} constructed a targeted dataset of tweets, and explored several types of potential features to build accurate predictive models based on machine learning to infer political leaning. Time series forecasting of political leaning is still booming recently. A recent work \cite{ng2022social} utilized time series forecasting to model the topic-specific daily volume of social media activities. The work in \cite{ng2023experimental} analyzed forecasting activity in several social media datasets, to capture different contexts occurring on multiple platforms such as Twitter and YouTube.

\begin{figure*}[htp]
  \centering
  \subfigure[Degree of Sentimentality in Twitter posts for each political leaning]
  {\includegraphics[scale=0.2]{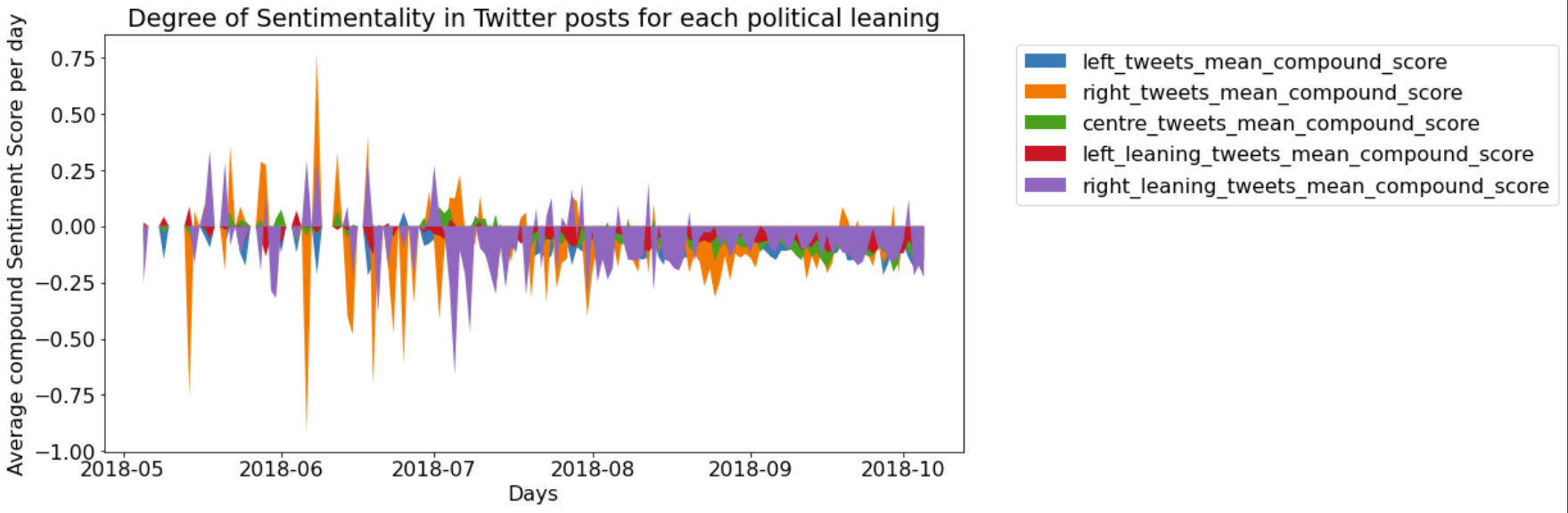}}\quad
  \subfigure[Degree of Sentimentality in GAB posts for each political leaning]
  {\includegraphics[scale=0.2]{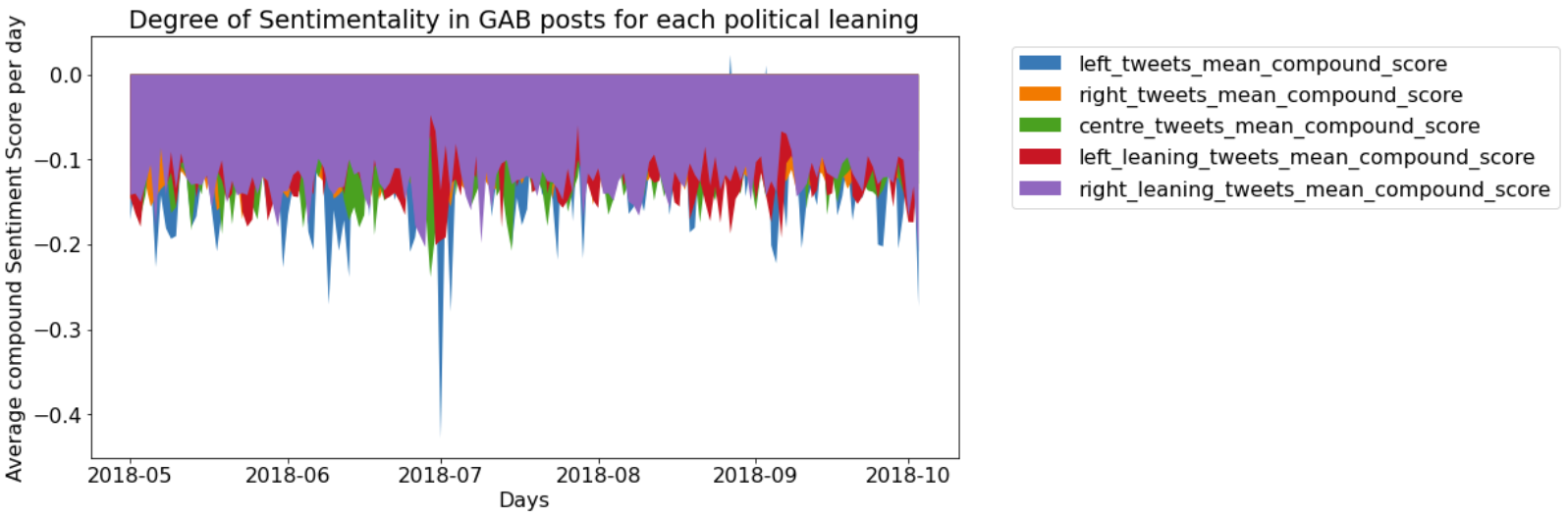}}
  \caption{Average Compound Sentiment score of Twitter and GAB posts per day}
  \label{fig:Sentiment}
\end{figure*}

\section{Datasets}

In this research, we utilized publicly available datasets from Twitter\cite{brena2019news} and Gab\cite{9680097}. The Twitter dataset\cite{brena2019news} consists of tweets that share news article URLs related to political topics from selected news media sources. The data spans from January 2018 to October 2018, comprising a total of 722,685 tweets. Our Gab data \cite{9680097}  comprises 1,345,279 posts from the same time span from January 2018 to October 2018. We have generated another comprehensive media bias dataset using web scraping tools from Allsides.com\footnote{\url{http://www.allsides.com/media- bias/ media-bias-rating-methods}}. AllSides utilizes user community for validation, assigns political leaning on a scale on news articles and media outlets.
We first analyze the sentiment polarity of posts that share news articles to understand properties of our datasets. We use VADER\cite{hutto2014vader} to obtain \emph{average compound score} of each post. It is evident from Figure \ref{fig:Sentiment}a, Twitter generally has negative sentiment polarity in posts that share right and right-leaning news articles while maintaining less negative value over other political leanings. Figure \ref{fig:Sentiment}b illustrates that the Gab dataset, has more frequent negative sentiment overall political leaning labels. It also shows that left posts have more negative sentiment over other political leanings. These analyses clearly indicates that two datasets have distinct political ideologies.

\section{Methodology}

%This section will provide a detailed explanation of the methodology used, starting with data preprocessing and extending to connect the models to our case. 
%It will also include the hyperparameters derived through hyperparameter tuning.

\subsection{Data preprocessing}

\par We label the Twitter and Gab posts to their corresponding political leaning using the political bias of collected political media bias outlets. That is if the news domain in a social media post has political leaning \emph{p}, we label the social media post as \emph{p}. In this study, we consider \emph{five} political leaning labels $p=\{left,left-leaning, center,right-leaning, right\}$. 

In this process, we also extracted the timestamps for each tweet or Gab post in two methods. (i) We calculated the respective political leaning post frequencies for each day, and (ii) we preprocessed frequencies of likes based on their respective political leaning for each day. Due to outliers and very small postings in other months we use posts from January to April in our experiments.
%the preprocessed data frames from January to May. Finally, we have data frames from May to October ready to be used for fitting models for each political leaning.

\subsection{Timeseries forecasting models.}
%Now that we have preprocessed the data, it was then the time to forecast the results of each political leaning. 
We used 5-time series forecasting models in this work. 
%So here is how we integrated the mathematical formulas into our case.
\subsubsection{ SARIMA Model}
%Initially, we used the Dickey-Fuller test for the preprocessed data and found that each political leaning data is not stationary.
As our Twitter and Gab datasets are non-stationary with political leaning, we choose the SARIMA model for forecasting. It is a statistical model that is a combination of the autoregression (where the value at the current time is forecasted in the linear combination of previous times until p), and moving average ( where past forecast errors are used in linear combination to forecast present time value) models with seasonality. So, the time series forecasting for a timestep $t$ is given by:
%Altogether the equation of the model that is used to forecast tweets, gab posts, gab post likes, and tweet likes for each leaning is:-
%Integrated( differencing factor to make it stationary d),

\begin{equation}
\begin{split}
    y_t = c &+ \sum_{n=1}^{p} \alpha_n y_{t-n} + \sum_{n=1}^{q} \theta_n \varepsilon_{t-n} \\
    &+ \sum_{n=1}^{P} \phi_n y_{t-sn} + \sum_{n=1}^{Q} \eta_n \varepsilon_{t-sn} + \varepsilon_t
\end{split}
\end{equation}

where $y_x$ is the frequency at time $x$, $c$ is the constant term, $\alpha_n$ is the autoregressive coefficient, $\theta_n$ is the moving average coefficient, $\phi_n$ is the seasonal autoregressive coefficient, $\eta_n$ is the seasonal moving average coefficient, $\varepsilon_t$ is the white noise error term. Also, the notations p, d, q, P, Q, and s are obtained by grid search on the SARIMAX function which varies for each political leaning in our preprocessed data. 
% y_{t-n} &\text{ is the frequency of either tweets, gab posts,} &\text{or tweet likes or gab likes at time } t-n, \\

%Here, in this equation, p,d,q, P, Q,s are found out by Grid search which varies for each political leaning preprocessed data, and their values are noted below. These values are initialized in the formula by the package statsmodels.tsa.statespace.sarimax which provides the SARIMAX function that automatically finds the parameter values that maximize the likelihood of observing the given time series under the SARIMA model. This is done using optimization algorithms like the Newton-Raphson method.

\subsubsection{LSTM Model}
Due to the limitation of SARIMA capturing very simple patterns and linear dependencies between variables we use two types of LSTM methods. One method takes only the previous day's data, while the other takes the past two week's data as input to make the next-day prediction. The LSTM methods used in this work are given below:
%We use two types of LSTM methods for both posts and likes time series data from both datasets.
% which can capture complex long-term dependencies

\begin{equation}
\begin{split}
i_t &= \sigma(W_i \cdot [h_{t-1}, x_t] + b_i) \\
f_t &= \sigma(W_f \cdot [h_{t-1}, x_t] + b_f) \\
g_t &= \tanh(W_g \cdot [h_{t-1}, x_t] + b_g) \\
c_t &= f_t \cdot c_{t-1} + i_t \cdot g_t \\
o_t &= \sigma(W_o \cdot [h_{t-1}, x_t] + b_o) \\
h_t &= o_t \cdot \tanh(c_t)
\end{split}
\end{equation}

where $i_t$ is the input gate activation at time step t, $f_t$ is the forget gate activation at time step t, $g_t$ is the cell state update at time step t, $c_t$ and $c_{t-1}$ are the cell state at time steps t and t-1 respectively, $o_t$ is the output gate activation at time step t, $h_t$ and $h_{t-1}$ are the hidden state at time steps t and t-1 respectively, $x_t$ is the input feature vector at time step t, $\sigma$ is the sigmoid activation function to squash the input values between 0 and 1, $\tanh$ is the hyperbolic tangent activation function to squash the input values between -1 and 1

The only difference that comes into play for our second LSTM is that we use a 14-day look back in place of single feature vector input $x_t$. Thus the input to the LSTM methods is a concatenation of all vectors $x_t = [x_t; x_{t-1}; x_{t-2}; \ldots; x_{t-13}]$. Other parameters like hidden states, epochs, and optimizers used in LSTM are set by hyperparameter tuning. We used the RMSE loss function in all our LSTM models.

%Once LSTMs are given a number of hidden layers they automatically handle the initialization and updating of weights and biases during the training process.
%Behind the scenes, during the training process, the LSTM layer updates its weights and biases using backpropagation and gradient descent to minimize the difference between the predicted outputs and the true labels. 
%But there are other parameters like epochs, optimizer, and loss function which play a major role in predicting with the above formulae working of LSTM. And these parameters are found by hyperparameter tuning which is a process of finding the best combination of these parameters to get less error. These values are obtained differently for some of the leaning and the same for the some of political leaning and we will observe them in the next sub-section.

%\vspace{10pt}
\subsubsection{Multistep time series forecasting model}
We modified the above LSTM model to make multi-step and beyond 1-day predictions for both posts and likes in each political leaning in both datasets. We mainly focus on making predictions for the next 5 days from the given 14-day look-back data.
% We have made a few changes to predict the next 7 sequences of days of political leaning posts or likes. These changes to the model are mainly in the output sequence. In the previous case, the output sequence was only one value at t but on applying this model we would be getting the output sequence of the next 7 days meaning by using this model we would get 7 next day predicted values.
%A change in Teacher forcing (a technique for training recurrent neural networks that substitute ground truth for model output from a previous time step) in the previous model resulted in this model. 
We utilized \emph{Teacher Forcing}~\cite{lamb2016professor} to achieve multistep forecasting to predict the entire output sequence  [t+1, t+2, ..., t+5] from the multistep look back sequence [t-13, t-12, ..., t-1, t]. In teacher forcing, instead of using LSTM's own generated output as input for the next time step, the ground truth or target sequence is used as input to the model at each time step during training. This leads to faster convergence of the forecasting model and a more stable training process.
%So, Teacher forcing is simple in a typical LSTM for single-step forecasting. Input sequences [t-13, t-12,..., t-1] and their associated single-step outputs [t] from the training data are presented to the LSTM during training. Using a loss function like Mean Squared Error (MSE), the LSTM is trained to reduce the difference between its predictions (for time step t) and the actual target values (ground truth). The LSTM creates a single prediction for the following time step t at each training step and contrasts it with the true value t derived from the training data. For each input-output pair in the training dataset, this process is repeated. But whereas in multistep time series forecasting the LSTM is asked to predict the entire output sequence [t+1, t+2, ..., t+7] given the input sequence [t-13, t-12, ..., t-1].
%And remaining every calculation is exactly the same as above.
The pipeline of our multistep time series forecasting model with teacher enforcing is :
\begin{enumerate}
    \item For each of the 5 prediction steps, we use a RNN or similar architecture. The model takes as input a sequence of 14 historical data points (lookback) and produces an output for the next day.
    
    \item During training, for each prediction step,  the true value for the corresponding day as part of the input sequence is provided. This enforces the model to learn accurate dependencies between historical and future data points.
    
    \item The loss function is the sum of losses for each prediction step, computed as the difference between predicted and true values. This encourages the model to refine its predictions iteratively.
    
    \item Once trained, the model can be deployed for forecasting by feeding it the most recent 14 days of data. It will generate predictions for the next 5 days, utilizing the learned temporal dependencies to make accurate forecasts.
\end{enumerate}

%\vspace{10pt}
\subsubsection{Gated Recurrent Unit.}
We used another RNN model \emph{GRU} which is considered simpler than LSTM and assists with capturing long-term dependencies in sequence data. The only difference is it combines the forget and input gates of LSTM into a single update gate and merges the cell state and hidden state of LSTM into a single hidden state. We use the GRU model as given below:

\begin{equation}
\begin{split}
  % Update gate (z_t) calculation
  z_t &= \sigma(W_z \cdot x_t + U_z \cdot h_{t-1} + b_z) \\
  % Reset gate (r_t) calculation
  r_t &= \sigma(W_r \cdot x_t + U_r \cdot h_{t-1} + b_r) \\
  % Candidate hidden state (\tilde{h}_t) calculation
  \tilde{h}_t &= \tanh(W_h \cdot x_t + U_h \cdot (r_t \odot h_{t-1}) + b_h) \\
  % Hidden state (h_t) calculation
  h_t &= (1 - z_t) \odot h_{t-1} + z_t \odot \tilde{h}_t
\end{split}
\end{equation}

where $z_t$ is the update gate at time step t, $r_t$ is the reset gate at t, $\tilde{h}_t$ is the candidate hidden state at t, $h_t$ is the hidden state at t, $x_t$ is the input feature vector at t, $h_{t-1}$ is the hidden state at the previous time step (t-1), $W_z, W_r, W_h$ are weight matrices of update gate, reset gate, and candidate hidden state respectively, associated with the current state, $U_z, U_r, U_h$ are Weight matrices for the update gate, reset gate, and candidate hidden state, associated with the previous state, $b_z, b_r, b_h$ are bias terms, and $\odot$ is the Element-wise multiplication (Hadamard product). We followed the same approach given for the LSTM methods to input a 14-day look back for the next timestep prediction.

%We used TensorFlow Keras to automate this model working and it initializes the weight matrices at the current and previous states, as well as the bias terms, for the GRU units. During training, these weights and biases are updated to optimize the model's performance on the given task. Basically, this is a step GRU but we have used 14 days look back so instead of $x_t$ we would get a vector of the past 14 days of either tweet post frequency, tweet likes, gab post frequency, or gab post likes, and input vector in each formula looks like:-
%\begin{align*}
%x_t = [x_t, x_{t-1}, x_{t-2}, \ldots, x_{t-13}]
%\end{align*}

\begin{table*}[htbp]
%\medium
\centering
\caption{RMSES of Time Series Forecasting of Tweets Frequencies from Twitter Dataset}
\label{TableI}
%\resizebox{\columnwidth}{!}{%
\renewcommand{\arraystretch}{1.25}
\begin{tabular}{|p{6.0cm}|c|c|c|c|c|}
\hline
\textbf{Models} & \textbf{Left} & \textbf{Right} & \textbf{Center} & \textbf{Left Leaning} & \textbf{Right Leaning} \\
\hline
SARIMA & \textbf{66.10} & \textbf{31.29} & \textbf{70.15} & \textbf{155.72} & \textbf{13.07} \\
\hline
LSTM (1 day feedback) & \begin{tabular}[c]{@{}l@{}}Train :- 16.76\\ Test :-  159.58\end{tabular} & \begin{tabular}[c]{@{}l@{}}Train :- 10.59\\ Test :- 63.73\end{tabular} & \begin{tabular}[c]{@{}l@{}}Train :- 32.65\\ Test :- 161.48\end{tabular} & \begin{tabular}[c]{@{}l@{}}Train :- 51.36\\ Test :- 369.79\end{tabular} & \begin{tabular}[c]{@{}l@{}}Train :- 6.28\\ Test :- 30.86\end{tabular} \\
\hline
LSTM (14 days feedback) & \begin{tabular}[c]{@{}l@{}}Train :- 17.74\\ Test :- 278.99\end{tabular} & \begin{tabular}[c]{@{}l@{}}Train :- 2.99\\ Test :- 105.28\end{tabular} & \begin{tabular}[c]{@{}l@{}}Train :- 18.11\\ Test :-  275.17\end{tabular} & \begin{tabular}[c]{@{}l@{}}Train :- 19.74\\ Test :-  774.52\end{tabular} & \begin{tabular}[c]{@{}l@{}}Train :- 1.88\\ Test :- 49.28\end{tabular} \\
\hline
GRU( 14 days feedback) & \begin{tabular}[c]{@{}l@{}}Train :- 51.97\\ Test :- 329.36\end{tabular} & \begin{tabular}[c]{@{}l@{}}Train :- 16.35\\ Test :- 437.32\end{tabular} & \begin{tabular}[c]{@{}l@{}}Train ;- 94.77\\ Test :- 184.6\end{tabular} & \begin{tabular}[c]{@{}l@{}}Train :- 428.02\\ Test :- 1002.4\end{tabular} & \begin{tabular}[c]{@{}l@{}}Train :- 19.82\\ Test :- 39.62\end{tabular} \\
\hline
\hline
Multistep Forecasting (14 days feedback and 5 next days predicting) & \begin{tabular}[c]{@{}l@{}}t+1 :- 147.15\\ t+2 :- 238.26\\ t+3 :- 285.23\\ t+4 :- 307.02\\ t+5 :- 310.57\end{tabular} & \begin{tabular}[c]{@{}l@{}}t+1 :- 50.29\\ t+2 :- 73.98\\ t+3 :- 89.28\\ t+4 :- 98.77\\ t+5 :- 100.92\end{tabular} & \begin{tabular}[c]{@{}l@{}}t+1 :- 131.08\\ t+2 :- 195.88\\ t+3 :- 236.04\\ t+4 :- 266.49\\ t+5 :- 280.76\end{tabular} & \begin{tabular}[c]{@{}l@{}}t+1 :- 279.15\\ t+2 :- 440.73\\ t+3 :- 541.36\\ t+4 :- 605.67\\ t+5 :- 648.48\end{tabular} & \begin{tabular}[c]{@{}l@{}}t+1 :- 23.05\\ t+2 :- 32.16\\ t+3 :- 38.54\\ t+4 :- 41.26\\ t+5 :- 42.15\end{tabular} \\
\hline

\end{tabular}%
\end{table*}

\begin{table*}[htbp]
%\medium
\centering
\caption{RMSES of Time Series Forecasting of Likes Frequencies from Twitter Dataset.}
\label{TableII}
%\resizebox{\columnwidth}{!}{%
\renewcommand{\arraystretch}{1.25}
\begin{tabular}{|p{6.0cm}|c|c|c|c|c|}
\hline
\textbf{Models} & \textbf{Left} & \textbf{Right} & \textbf{Center} & \textbf{Left Leaning} & \textbf{Right Leaning} \\
\hline
SARIMA & 216.48 & 55.95 & 228.02 & \textbf{1364.16} & 25.76 \\
\hline

LSTM (1 day feedback) & \begin{tabular}[c]{@{}l@{}}Train :- 185.22\\ Test :-  351.86\end{tabular} & \begin{tabular}[c]{@{}l@{}}Train :- 13.95\\ Test :- 95.63\end{tabular} & \begin{tabular}[c]{@{}l@{}}Train :- 69.23\\ Test :- 529.16\end{tabular} & \begin{tabular}[c]{@{}l@{}}Train :- 669.69\\ Test :- 5312.20\end{tabular} & \begin{tabular}[c]{@{}l@{}}Train :- 8.30\\ Test :- 62.22\end{tabular} \\
\hline

LSTM (14 days feedback) & \begin{tabular}[c]{@{}l@{}}Train :- 193.23\\ Test :- 285.68\end{tabular} & \begin{tabular}[c]{@{}l@{}}Train :- 33.80\\ Test :- 169.71\end{tabular} & \begin{tabular}[c]{@{}l@{}}Train :- 61.05\\ Test :-  642.67\end{tabular} & \begin{tabular}[c]{@{}l@{}}Train :- 175.26\\ Test :-  3647.8\end{tabular} & \begin{tabular}[c]{@{}l@{}}Train :- 2.76\\ Test :- 51.17\end{tabular} \\
\hline

GRU( 14 days feedback) & \begin{tabular}[c]{@{}l@{}}Train :- 217.69\\ Test :- \textbf{116.7}\end{tabular} & \begin{tabular}[c]{@{}l@{}}Train :- 67.65\\ Test :- \textbf{28.82}\end{tabular} & \begin{tabular}[c]{@{}l@{}}Train ;- 167.86\\ Test :- \textbf{197.32}\end{tabular} & \begin{tabular}[c]{@{}l@{}}Train :- 740.97\\ Test :- 4259.36\end{tabular} & \begin{tabular}[c]{@{}l@{}}Train :- 26.71\\ Test :- \textbf{7.59}\end{tabular} \\
\hline
\hline
Multistep Forecasting (14 days feedback and 5 next days predicting) & \begin{tabular}[c]{@{}l@{}}t+1 :- 389.58\\ t+2 :- 352.89\\ t+3 :- 341.11\\ t+4 :- 333.76\\ t+5 :- 403.50\end{tabular} & \begin{tabular}[c]{@{}l@{}}t+1 :- 123.28\\ t+2 :- 141.52\\ t+3 :- 139.45\\ t+4 :- 132.35\\ t+5 :- 144.41\end{tabular} & \begin{tabular}[c]{@{}l@{}}t+1 :- 546.24\\ t+2 :- 603.19\\ t+3 :- 564.96\\ t+4 :- 617.59\\ t+5 :- 573.48\end{tabular} & \begin{tabular}[c]{@{}l@{}}t+1 :- 1626.40\\ t+2 :- 1617.93\\ t+3 :- 1620.36\\ t+4 :- 1629.02\\ t+5 :- 3425.94\end{tabular} & \begin{tabular}[c]{@{}l@{}}t+1 :- 60.57\\ t+2 :- 56.32\\ t+3 :- 53.06\\ t+4 :- 60.41\\ t+5 :- 50.92\end{tabular} \\
\hline

\end{tabular}%
\end{table*}

\section{Experiments and Results}

\noindent \textbf{Hyper parameters Tuning.} We first list hyperparameters of our models and then discuss results. Since we have time series for each political leaning, we use an exclusive model for each political leaning in our experiments.
%As building these models from scratch is very tedious we used the statsmodels package for automating the SARIMA model, Keras for both LSTM models and the Multistep time series forecasting model, and tensorFlow packages for the GRU model. These packages usually provide functions that have the lines of code inbuilt which replicates the above-mentioned mathematics of these models. But there are some parameters that need to be set some values before training. The mathematics of the model will be working during the training phase but before that, we need to set these values to check which combination of these values is providing less error, and setting these values by trial and error is not advisable because there will be many combinations. So we hyper-tune these parameters. This technique will compute all the combinations and give out the best combination that produces less error.

\subsubsection{SARIMA Model hyperparameters.}
We used Grid search Hyperparameter tuning to get order parameters $p,d,q$, and seasonal order parameters $P, D, Q, S$ in SARIMA for both Twitter and Gab datasets. For Twitter posts frequency we found order parameters $(9,0,10)$ and seasonal order parameters $((2,1,1,12))$ to be optimal. Whereas for Twitter like we observed order=$(11,1,3)$, seasonal order=$(3,1,3,12)$ to be optimal for all political leanings.
%and for each learning the values are:\\
%Twitter Posts SARIMA model hyperparameters:
%\begin{itemize}
%    \item Left :- order = (9,0,10), seasonal order = (2,1,1,12)
%    \item Right :- order = (7,0,10), seasonal order = (2,1,1,12)
%    \item Centre :- order = (9,0,10), seasonal order = (2,1,1,12)
%    \item Left Leaned :- order = (9,0,10), seasonal order = (2,1,1,12)
%    \item Right Leaned :- order = (9,0,10), seasonal order = (2,1,1,12)
%\end{itemize}

%For Twitter likes we observed order=(11,1,3), seasonal order=(3,1,3,12) to be same for all political leanings.

Unlike Twitter, we noted that we get different optimal parameter values for Gab. For Gab posts frequency forecasting we set the following parameters.
\begin{itemize}
    \item Left  :- order=(7,1,10), seasonal order=(3,1,1,14)
    \item Right  :- order=(6,2,10), seasonal order=(4,1,1,11)
    \item Centre  :- order=(11,1,10), seasonal order=(2,1,1,14)
    %\item Left-leaning  :- order=(7,1,10), seasonal order=(3,1,1,14)
    %\item Right-leaning  :- order=(6,2,10), seasonal order=(4,1,1,14)
\end{itemize}

And, we use the following parameters for Gab likes frequency forecasting.
\begin{itemize}
    \item left  :- order=(11,1,6), seasonal order=(3,0,4,12)
    \item right  :- order=(9,1,11), seasonal order=(1,1,3,12)
    \item centre  :- order=(8,1,11), seasonal order=(4,0,0,12)
    %\item left-leaning  :- order=(11,1,6), seasonal order=(3,0,4,12)
    %\item right-leaning  :- order=(9,1,11), seasonal order=(1,1,3,12)
\end{itemize}

%\vspace{5pt}

%We use Bayesian Optimization to find Optimal Hyperparameters for deep-learning-based models. So LSTM(1-day lookback), LSTM( 14 days lookback), GRU(14 days lookback), and Multistep time series forecasting models hyperparameters are tuned using Bayesian Optimization. So the values are: \\

\subsubsection{LSTM Hyperparameters.} We use 4 hidden layers, trained with 100 epochs, and \emph{RMSProp} optimizer as hyperparameters for all political leaning forecasting in the Twitter dataset for both 1-day lookback and 14-day lookback. The same set of hyperparameters is used for both tweets frequency and likes frequency forecasting. Similarly, we used 4 hidden layers, 200 training epochs, and \emph{RMSProp} as optimizers for all experiments with the Gab data. We use \emph{Mean Squared Error} (MSE) loss in all LSTM experiments.
%to SARIMA, we used a different set of hyperparameters, as given below, for each political leaning in the Gab data. However, we followed the same set of parameters for all experiments in Gab.
%LSTM (1-day lookback) for Tweets:\\
%\begin{itemize}
%    \item Left :-  hidden layers = 4, optimizer = 'adam' , epochs =100
%    \item Right :- hidden layers = 4, optimizer = 'RMSProp', epochs = 96
%\end{itemize}

%LSTM for Centre, Left leaned, and Right leaned tweets have the same hyperparameters as Right tweets.
%Now for Tweet likes LSTM(1-day lookback) for all political leaning had the same hyperparameters as their respective leaning hyperparameters for LSTM of tweets.\\

%LSTM (1-day lookback) for Gab:\\
%\begin{itemize}
 %   \item Left  :-  hidden layers =4, epochs =200, optimizer = adam
  %  \item Right  :-  optimizer = RMSProp
   % \item Centre  :-  optimizer= RMSProp, epochs = 200, remaining is same.
    %\item Lean Right :-  optimizer = RMSProp, epochs =100 and remaining is same.
   % \item Lean Left :-   epochs =200, remaining is same.
        
%\end{itemize}

%Similar to Tweets Gab post likes models also have the same Hyper-parameter values.

%From above there is no difference between LSTM(1-day lookback), and LSTM(14 days lookback) except for the input vector. So every operation is the same therefore similar values of LSTM(1-day lookback) were also noticed in LSTM (14 days lookback).\\
%Finally, there are 2 more Hyperparameters that are common in all of these LSTMS that are batch size and loss function whose values are 1, mean squared error.
%\vspace{5pt}

\subsubsection{Multistep Time Series Forecasting Hyperparameters.}
Hyperparameters for multistep time series forecasting models differ only with the number of epochs we used in training. Other than that we use \emph{RMSProp} optimizer, 8 hidden layers with 8 hidden neurons in each layer, and MSE loss function as hyperparameters. We use 125 epochs for tweets, 150 epochs for Gab posts, and 100 epochs for likes data in general. 
%the following epochs in all experiments:
%\begin{itemize}
%    \item tweets forecasting:- 125
%    \item tweets likes forecasting:- 100
%    \item Gab posts forecasting:- 150
%    \item Gab likes forecasting:-  100
   % \item Left leaned :-   epochs =150, remaining same.
%\end{itemize}

%Hyperparameters for Tweet likes:\\
%\begin{itemize}
%    \item Left  :-  epochs =250, hidden neurons =8, batch =1, optimizer ="RMSProp".
%    \item Right:-  epochs =100, remaining all are same.
%    \item Centre:- epochs= 100, remaining all are same.
%    \item Left leaned :-  same as above.
%    \item Right leaned :- same as above.
%\end{itemize}

%Hyperparameters for GAB posts:\\
%\begin{itemize}
%    \item Left :- epochs=200, batch =1, optimizer ="RMSProp", hidden neurons=1.
%    \item Centre :- epochs = 75, remaining same.
%    \item Right :- epochs =150, remaining same.
%    \item Right leaned :- epochs =150, remaining same.
%    \item Left leaned :- epochs =200, remaining same.
%\end{itemize}

%Hyperparameters for GAB post likes:\\
%\begin{itemize}
%    \item Left  :-  hidden layers =8, epochs =150, batch =1, optimizer ="RMSProp".
%    \item Right  :- epochs =100, remaining are same.
%    \item Centre  :- epochs =100,remaining are same.
%    \item Left leaned  :-  epochs =175, remaining are same.
%    \item Right leaned  :-  epochs =100, remaining are same.
%\end{itemize}
%\vspace{3pt}
%All the above Multi-step time series forecasting models have a common loss function which is a mean squared error.
%\vspace{10pt}
\subsubsection{GRU Hyperparameters.}
We set dropout as 0.2, adam optimizer, \emph{MSE} loss, 100 training epochs, and a batch size of 16 for GRU forecasting models in both Twitter and Gab.
%Hyperparameters for tweets:\\
%\begin{itemize}
%    \item Left :- dropout= 0.2, optimizer=adam, loss = mse, epocs =100, batch =16.
%\end{itemize}
%remaining all leanings of tweets have the same parameter values.
%And here it appears to be the same for all leanings of the GRUs of GAB posts and likes as well.

\begin{table*}[htbp]
%\medium
\centering
\caption{RMSES of Time Series Forecasting of Posts frequencies in Gab Dataset.}
\label{TableIII}
%\resizebox{\columnwidth}{!}{%
\renewcommand{\arraystretch}{1.25}
\begin{tabular}{|p{6.0cm}|c|c|c|c|c|}
\hline
\textbf{Models} & \textbf{Left} & \textbf{Right} & \textbf{Center} & \textbf{Left Leaning} & \textbf{Right Leaning} \\
\hline
SARIMA & \textbf{37.04} & \textbf{263.40} & \textbf{78.12} & 63.90 & \textbf{98.66} \\
\hline
LSTM (1 day feedback) & \begin{tabular}[c]{@{}l@{}}Train :- 31.98\\ Test :-  48.23\end{tabular} & \begin{tabular}[c]{@{}l@{}}Train :- 247.32\\ Test :- 532.17\end{tabular} & \begin{tabular}[c]{@{}l@{}}Train :- 73.95\\ Test :- 164.32\end{tabular} & \begin{tabular}[c]{@{}l@{}}Train :- 57.76\\ Test :- 73.42\end{tabular} & \begin{tabular}[c]{@{}l@{}}Train :- 97.55\\ Test :- 142.31\end{tabular} \\
\hline
LSTM (14 days feedback) & \begin{tabular}[c]{@{}l@{}}Train :- 26.71\\ Test :- 46.64\end{tabular} & \begin{tabular}[c]{@{}l@{}}Train :- 219.08\\ Test :- 445.41\end{tabular} & \begin{tabular}[c]{@{}l@{}}Train :- 58.43\\ Test :-  106.05\end{tabular} & \begin{tabular}[c]{@{}l@{}}Train :- 49.76\\ Test :-  \textbf{57.04}\end{tabular} & \begin{tabular}[c]{@{}l@{}}Train :- 80.62\\ Test :- \textbf{97.86}\end{tabular} \\
\hline
GRU( 14 days feedback) & \begin{tabular}[c]{@{}l@{}}Train :- 38.68\\ Test :- 46.62\end{tabular} & \begin{tabular}[c]{@{}l@{}}Train :- 556.54\\ Test :- 799.49\end{tabular} & \begin{tabular}[c]{@{}l@{}}Train ;- 98.43\\ Test :- 176.40\end{tabular} & \begin{tabular}[c]{@{}l@{}}Train :- 69.3\\ Test :- 84.67\end{tabular} & \begin{tabular}[c]{@{}l@{}}Train :- 116.62\\ Test :- 165.96\end{tabular} \\
\hline
\hline
Multistep Forecasting (14 days feedback and 5 next days predicting) & \begin{tabular}[c]{@{}l@{}}t+1 :- 50.06\\ t+2 :- 66.83\\ t+3 :- 72.61\\ t+4 :- 69.55\\ t+5 :- 71.95\end{tabular} & \begin{tabular}[c]{@{}l@{}}t+1 :- 526.08\\ t+2 :- 691.85\\ t+3 :- 687.51\\ t+4 :- 744.08\\ t+5 :- 849.04\end{tabular} & \begin{tabular}[c]{@{}l@{}}t+1 :- 155.51\\ t+2 :- 201.90\\ t+3 :- 227.51\\ t+4 :- 225.89\\ t+5 :- 207.46\end{tabular} & \begin{tabular}[c]{@{}l@{}}t+1 :- 72.47\\ t+2 :- 89.76\\ t+3 :- 88.57\\ t+4 :- 79.95\\ t+5 :- 88.27\end{tabular} & \begin{tabular}[c]{@{}l@{}}t+1 :- 163.21\\ t+2 :- 225.88\\ t+3 :- 248.94\\ t+4 :- 218.84\\ t+5 :- 190.78\end{tabular} \\
\hline

\end{tabular}%
\end{table*}

\begin{table*}[htbp]
%\medium
\centering
\caption{RMSES of Time Series Forecasting of Likes of posts from Gab Dataset.}
\label{TableIV}
%\resizebox{\columnwidth}{!}{%
\renewcommand{\arraystretch}{1.25}
\begin{tabular}{|p{6.0cm}|c|c|c|c|c|}
\hline
\textbf{Models} & \textbf{Left} & \textbf{Right} & \textbf{Center} & \textbf{Left Leaning} & \textbf{Right Leaning} \\
\hline
SARIMA & 223.06 & \textbf{1265.42} & \textbf{310.95} & \textbf{235.66} &  \textbf{395.36}\\
\hline

LSTM (1 day feedback) & \begin{tabular}[c]{@{}l@{}}Train :- 239.56\\ Test :-  \textbf{211.34}\end{tabular} & \begin{tabular}[c]{@{}l@{}}Train :- 1217.28\\ Test :- 2549.69\end{tabular} & \begin{tabular}[c]{@{}l@{}}Train :- 256.88\\ Test :- 633.05\end{tabular} & \begin{tabular}[c]{@{}l@{}}Train :- 233.39\\ Test :- 256.56\end{tabular} & \begin{tabular}[c]{@{}l@{}}Train :- 357.77\\ Test :- 480.48\end{tabular} \\
\hline

LSTM (14 days feedback) & \begin{tabular}[c]{@{}l@{}}Train :- 231.28\\ Test :- 227.59\end{tabular} & \begin{tabular}[c]{@{}l@{}}Train :- 919.72\\ Test :- 2389.58\end{tabular} & \begin{tabular}[c]{@{}l@{}}Train :- 251.63\\ Test :-  694.82\end{tabular} & \begin{tabular}[c]{@{}l@{}}Train :- 208.62\\ Test :-  294.13\end{tabular} & \begin{tabular}[c]{@{}l@{}}Train :- 317.45\\ Test :- 523.94\end{tabular} \\
\hline

GRU( 14 days feedback) & \begin{tabular}[c]{@{}l@{}}Train :- 251.4\\ Test :- \textbf{212.8}\end{tabular} & \begin{tabular}[c]{@{}l@{}}Train :- 1831.7\\ Test :- 2208.50\end{tabular} & \begin{tabular}[c]{@{}l@{}}Train ;- 277.15\\ Test :- 630.07\end{tabular} & \begin{tabular}[c]{@{}l@{}}Train :- 247.06\\ Test :- 272.13\end{tabular} & \begin{tabular}[c]{@{}l@{}}Train :- 390.66\\ Test :- 503.55\end{tabular} \\
\hline
\hline

Multistep Forecasting (14 days feedback and 5 next days predicting) & \begin{tabular}[c]{@{}l@{}}t+1 :- 280.30\\ t+2 :- 277.96\\ t+3 :- 327.75\\ t+4 :- 318.86\\ t+5 :- 314.24\end{tabular} & \begin{tabular}[c]{@{}l@{}}t+1 :- 2547.45\\ t+2 :- 33683.13\\ t+3 :- 3381.94\\ t+4 :- 3250.28\\ t+5 :- 3343.77\end{tabular} & \begin{tabular}[c]{@{}l@{}}t+1 :- 592.32\\ t+2 :- 720.73\\ t+3 :- 852.81\\ t+4 :- 774.91\\ t+5 :- 850.64\end{tabular} & \begin{tabular}[c]{@{}l@{}}t+1 :- 318.20\\ t+2 :- 336.51\\ t+3 :- 333.83\\ t+4 :- 320.60\\ t+5 :- 345.58\end{tabular} & \begin{tabular}[c]{@{}l@{}}t+1 :- 610.65\\ t+2 :- 705.73\\ t+3 :- 724.45\\ t+4 :- 703.91\\ t+5 :- 773.96\end{tabular} \\
\hline

\end{tabular}%
\end{table*}

%\subsection{Insights of Datasets.}

%As there are 2 datasets of Twitter and GAB, the first and foremost thing done here was diving deep into the exploration of insights of datasets. Since the above 2 social media APIs have timestamps for the creation of posts. These timestamps are utilized to analyze the statistics of postings and like counts per day or per month. First, the trends of posts and their respective likes in each of the dataset per day has been explored and found that they are non-stationary time series graphs. Next, the exploration of statistics of posts and likes was changed to observe per month, and found some interesting multi-line plots of posts and likes for 2 datasets. These multi-line plots are obtained by taking the average frequency of posts per month on the y-axis and months on the x-axis. But the way of obtaining multi-line plots for likes is different because their total likes have been calculated per month instead of average, mode, and median. This is to analyze the statistical behavior of every political leaning and can be distinguished in the graph up to most of the extent only by taking the sum of likes per month. \\[2 pt]

%\subsection{Analyzing Degree of Sentimentality.}

%\section{Tables}

\noindent \textbf{Results of timeseries forecasting models.}
%We use the above-mentioned time-series forecasting models with corresponding hyperparameters on both Twitter and Gab datasets. 
In this section, we give forecasting results for both post frequencies and likes frequencies. We also give results for both next-day predictions and t+5 days predictions using the LSTM model in all experiments. We use the split of 70\% training and 30\% test for all models. One can see test rmse values higher than train rmse values and can easily come to a conclusion that model might be overfitted. But it is not the case because with severe experimentation we had given a sufficient training size. The reason for higher test rmse values is that we have applied forecasting on non-stationary data and this non-stationarity can be made to stationary and then applied to training for better rmses but non-stationarity isn't changed for a reason and the reason is to observe trends in the social media posts.  
One observation from Tables I, II, III, and IV is that our LSTM model with a 14-day look back gives optimal forecasting results with training instances in all experiments. However, the same model underperforms with different test instances. We consider only test results in all our below analysis. Although SARIMA performs better in forecasting tasks overall, it is interesting from our results that some political leaning for the same tasks in the same dataset fits well with other models.
%This part covers the experimentation of Time Series forecasting for Twitter posts, Twitter likes and GAB posts and gab likes. 5-time series forecasting models were tested by utilizing the time stamp of posts from the raw data. All these models were trained using the above-mentioned parameters and the results that were obtained are tabulated as below in the Tables section.

Table I presents the RMSE of all models to forecast tweet frequencies in the Twitter dataset. It is evident that SARIMA outperforms 2x times other models for next-day forecasting in all political leanings. In terms of multistep forecasting with LSTM, we notice that our model can forecast long-term projections in the 'Center' time series. Also, we see an exponential rise in RMSE for 'Left' and 'Left Leaning' labels for short-term predictions which gets smooth for long-term forecasting. Table II gives RMSE of all models to forecast likes frequencies in tweets. Here, the GRU model with a 14-day lookback is able to outperform all models, including SARIMA, by 2x times. Surprisingly, all models give high RMSE to forecast likes frequencies from 'Left Leaning' and SARIMA is the only model to give moderately lower RMSE. The analysis of multistep forecasting resembles that of results from Table I. It is also notable from Tables I and II that the RMSE of 'Right' and 'Right Leaning' is comparatively lower than other political leaning for the Twitter dataset. This is because less amount of data in these labels in Twitter data.

We notice from Tables III and IV that SARIMA model forecasts both posts frequencies and likes frequencies in Gab data by about 1.5x times. We also notice that the RMSE of 'Right' timeseries are high in both Gab experiments. This can arise because of high range of data in this label. Unlike Twitter, we do not find any patterns emerging from likes frequencies forecasting with Gab data in Table IV. However, Table III suggests that the proposed model with \emph{Teacher Forcing} can assist with long-term forecasting of 'Center' and 'Right Leaning' post frequencies in Gab. It is evident from Tables III and IV that the RMSE of 'Left' and 'Left Leaning' is comparatively lower than other political leaning for Gab dataset due to insufficient data.

\section{Conclusion}
In this work, our effort to forecast political bias in online social media has contributed valuable insights to the understanding of temporal dynamics in political conversations. In this paper, a novel method that is different from previously related work aims at time series forecasting of politically leaned social media posts and their likes from Twitter and Gab. We analyzed multiple forecasting models to predict both the next time step and future steps. In this work, we note that existing time series models are capable of forecasting political bias in online social media activities. However, we also note that these models are sensitive and their performance drops for the large quantity of input data. The future directions can be innovating novel time series models specific for political bias problems that can handle shortcomings of existing models. 

%5 Time series forecasting models named SARIMA, LSTM(1 day lookback), LSTM(14 days lookback) , Multistep forecasting( 14 days lookback and 7 next days predicting) and GRU( 14 days lookback) have been implemented for 2 datasets. Based on RMSES of forecasting, the model with LSTM (14 days lookback) seems to be performing better than other models in most of the cases on trained data.

\bibliographystyle{IEEEtran}
\bibliography{bibfile}

\end{document}